\documentclass[prd,showpacs]{revtex4}
\usepackage{graphics}
\def \beq{\begin{equation}}
\def \eeq{\end{equation}}
\def \beqa{\begin{eqnarray}}
\def \eeqa{\end{eqnarray}}

\begin{document}

\title{The continuum limit of quark number susceptibilities}
\author{Rajiv V.\ Gavai}
\email{gavai@tifr.res.in}
\author{Sourendu Gupta}
\email{sgupta@tifr.res.in}
\affiliation{Department of Theoretical Physics, Tata Institute of Fundamental
         Research,\\ Homi Bhabha Road, Mumbai 400005, India.}

\begin{abstract}
We report the continuum limit of quark number susceptibilities in quenched
QCD. Deviations from ideal gas behaviour at temperature $T$ increase
as the lattice spacing is decreased from $T/4$ to $T/6$, but a further
decrease seems to have very little effect. The measured susceptibilities
are 20\% lower than the ideal gas values, and also 10\% below the hard
thermal loop (HTL) results. The off-diagonal susceptibility is several
orders of magnitude smaller than the HTL results. We verify a strong
correlation between the lowest screening mass and the susceptibility.
We also show that the quark number susceptibilities give a reasonable
account of the Wroblewski parameter, which measures the strangeness
yield in a heavy-ion collision.
\end{abstract}
\preprint{TIFR/TH/02-05, hep-lat/0202006}
\maketitle

\section{Introduction}

With the RHIC now in its second year of running, it is important to
pin down lattice predictions for the high temperature phase of QCD.
In this context, fully non-perturbative measurements of quark number
susceptibilities \cite{gott,gavai,qnch,2fl} are important for four
reasons. Firstly, they are directly related to experimental measurements
of event-to-event fluctuations in particle production \cite{koch}.
Secondly, earlier results \cite{qnch,2fl} showed a strong jump in the
susceptibility across the phase transition, but indicated a significant
departure from weak-coupling behaviour, and it is important to check
whether this persists into the continuum. Third, resummed perturbative
computations of this quantity have now become available \cite{toni}, thus
making it possible to accurately test the importance of non-perturbative
contributions to this quantity. Finally, with continuum extrapolated results
in hand, one can address the question of whether the strangeness production
seen in heavy-ion collisions can be quantitatively explained as a signal
of the quark-gluon plasma.

We have recently presented systematic results for quark number
susceptibilities in quenched QCD \cite{qnch} as well as for QCD with two
flavours of light dynamical quarks \cite{2fl}.  In these studies a large
range of temperatures, $T$, was covered at a series of different quark
masses, $m$.  These computations were done at a fixed cutoff with the
lattice spacing, $a=T/4$ while keeping the finite volume effects under
control so that the thermodynamic limit could be taken reliably.  There
was a 3--5\% difference between the quenched and dynamical computations.
In this paper we examine the cutoff dependence of the quenched results
and report their continuum (zero lattice spacing) limit.

The partition function of QCD with three quark flavours is
\beq
   Z(T,\mu_u,\mu_d,\mu_s) = \int{\cal D}U \exp[-S(T)]
       \prod_{f=u,d,s}\det M(T,m_f,\mu_f),
\label{part}\eeq
where the temperature $T$ determines the size of the Euclidean time
direction, $S(T)$ is the gluonic part of the action and the determinants
of the Dirac matrices, $M$, contain as parameters the quark masses, $m_f$,
and the chemical potentials $\mu_f$ for each of the flavours $f=u,d,s$.
We also define the chemical potentials $\mu_0=\mu_u+\mu_d+\mu_s$,
$\mu_3=\mu_u-\mu_d$ and $\mu_8=\mu_u+\mu_d-2\mu_s$, which correspond to
the diagonal flavour $SU(3)$ generators. Note that $\mu_0$ is the usual
baryon chemical potential and $\mu_3$ is an isovector chemical potential.

Quark number densities are defined as
\beq
   n_i(T,\mu_u,\mu_d,\mu_s) = {T \over V} \frac{\partial\log Z}{\partial\mu_i},
\label{numb}\eeq
and the susceptibilities as
\beq
   \chi_{ij}(T,\mu_u,\mu_d,\mu_s) =
    {T \over V}  \frac{\partial^2\log Z}{\partial\mu_i\partial\mu_j},
\label{susc}\eeq
where $V$ denotes the spatial volume. The subscripts $i$ and $j$ are
either of the index sets $f$ or $\alpha=0$, 3 and 8.  We lighten the
notation by writing the diagonal susceptibilities $\chi_{ii}$ as $\chi_i$.
We determine the susceptibilities at zero chemical potential, $\mu_f=0$.
In this limit, all $n_i(T)=0$. Since we work with $m_u=m_d=m<m_s$,
we also have $\chi_{03}=0$.

Flavour off-diagonal susceptibilities such as
\beq
   \chi_{ud} = \left(\frac{T}{V_3}\right)
         \left\langle{\rm tr} M_u^{-1} M_u'\;
                     {\rm tr} M_d^{-1} M_d'\right\rangle
\label{offd}\eeq
are given entirely in terms of the expectation values of disconnected
loops. Since $m_u=m_d$, we obtain
$\chi_{us}=\chi_{ds}$ with each defined by an obvious generalization
of the formula above. Of the flavour diagonal susceptibilities we shall use
\beq
   \chi_s = \left(\frac{T}{V_3}\right) \left[
         \left\langle\left({\rm tr} M_s^{-1} M_s'\right)^2
                     \right\rangle +
         \left\langle{\rm tr}
             \left(M_s^{-1} M_s'' - M_s^{-1}M_s'M_s^{-1}M_s'\right)
                  \right\rangle\right].
\label{chis}\eeq
$\chi_u=\chi_d$ are given by a generalization of this formula.
Numerically, the simplest quantity to evaluate is the diagonal iso-vector 
susceptibility
\beq
   \chi_3 = \frac12 \left(\frac{T}{V_3}\right) 
         \left\langle{\rm tr} \left(M_u^{-1} M_u'' 
                 - M_u^{-1}M_u'M_u^{-1}M_u'\right)\right\rangle.
\label{chi3}\eeq
Two more susceptibilities are of interest. These are the baryon number and 
electric charge susceptibilities, 
\beq
   \chi_0 = \frac19\left(4\chi_3+\chi_s+4\chi_{ud}+4\chi_{us}\right)
       \qquad{\rm and}\qquad
   \chi_q = \frac19\left(10\chi_3+\chi_s+\chi_{ud}-2\chi_{us}\right).
\label{chi0q}\eeq
Note that $\chi_0$ is the baryon number susceptibility for three flavours
of quarks. As a result, this expression differs from the iso-singlet
quark number susceptibility for two flavours, defined in \cite{gott},
both in overall normalization and by terms containing strangeness.
In our numerical work we have chosen to use staggered quarks. Hence, to
normalise to one flavour of continuum quarks and compensate for fermion
doubling on the lattice, we have to multiply each of the traces in eqs.\
(\ref{offd}--\ref{chi3}) by a factor of $1/4$.

The quark mass appears in two logically distinct places---
first in the operators which define the susceptibilities in eqs.\
(\ref{offd}--\ref{chi3}), and secondly in the determinant in the
partition function of eq.\ (\ref{part}) which defines the weight for
the averaging of these operators. The first defines a valence quark mass
and the second the sea quark mass. In this paper we adopt the quenched
approximation, whereby the determinants in eq.\ (\ref{part}) are set
equal to unity. However, we shall vary the valence quark masses. As
shown earlier, unquenching the sea quarks changes results by 3--5\%
\cite{qnch,2fl}.

Note that all the flavour off-diagonal susceptibilities are exactly zero
in an ideal gas. We shall denote by $\chi^3_{FFT}$ the ideal gas value
for $\chi_3$.  On an $N_t\times N_s^3$ lattice with $N_c$ colours and
lattice spacing $a$, we find
\beq
   a^2\chi^3_{FFT} = \frac{N_c}{2N_tN_s^3}\sum_p \left\{ 
      D^{-2} \sin^2p_0\cos^2p_0+ D^{-1}\left(\sin^2p_0-\frac12\right)\right\},
\label{fft}\eeq
where the spectrum of momenta is $p_0=(2\pi/N_t)(n+1/2)$
with $0\le n<N_t$ and $p_i=2\pi n/N_s$ with $0\le n<N_s$, and
$D=(ma)^2+\sum_\mu\sin^2p_\mu$.  For a given $m/T_c$ and $T/T_c$, the
value of the quark mass in lattice units is
\beq
   ma=\left(\frac1{N_t}\right)\,
           \left(\frac m{T_c}\right)\,\left(\frac{T_c}T\right).
\label{mass}\eeq
The formula in eq.\ (\ref{fft}) is normalised such that there is exactly
one copy of each flavour of quarks. The values of $\chi_s$, $\chi_0$
and $\chi_q$ for an ideal gas can then be simply obtained as linear
combinations of $\chi^3_{FFT}$ for different valence quark masses.

\section{Lattice results}

\begin{figure}[hbt]\begin{center}
   \includegraphics{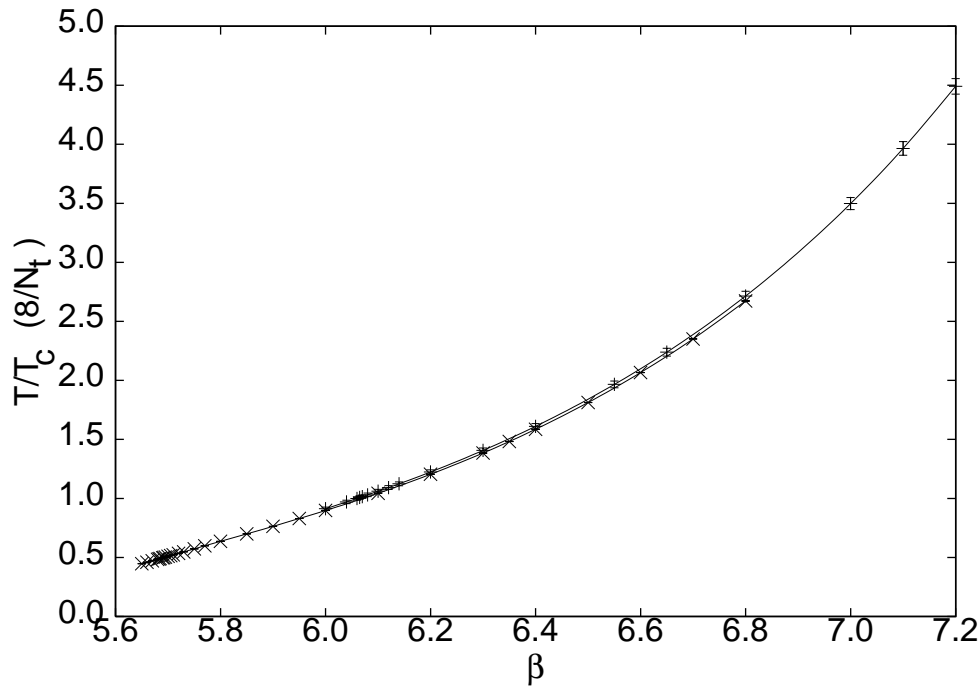}
   \end{center}
   \caption{The temperature for any value of $N_t$ for a given coupling
       $\beta$. Crosses denote data for $N_t=4$ lattices and pluses for
       $N_t=8$ lattices. The scale has been set by a corrected QCD 2-loop
       formula \cite{myold}. The difference between the two sets of data
       is a measure of scale breaking by residual power-law corrections
       and is negligible within statistical errors.}
\label{fg.temp}\end{figure}

We have previously reported measurements of $\chi_3$ in quenched QCD with
4 time slices ($N_t=4$) at temperatures of $1.1T_c$, $1.25T_c$, $1.5T_c$,
$2T_c$ and $3T_c$ \cite{gavai}. We have extended these measurements to
higher temperatures through two simulations on $4\times20^3$ lattices at
$\beta=6.6$ and $6.7$. These correspond to temperatures of $4.13T_c$ and
$4.7T_c$. All these computations are performed at fixed bare quark masses
of $m/T_c=0.03$, $0.3$, $0.5$, $0.75$ and 1 (the data for $m/T_c=0.5$
are new).

The continuum limit has been taken by going to several smaller lattice
spacings. One set of computations is performed with a lattice spacing
$a=1/6T$ which is 33\% smaller than the lattice spacing used earlier.
We have taken data on $6\times20^3$ lattices at $\beta=6.0625$, $6.3384$
and $6.7$ corresponding to $T/T_c=1.333$, 2 and $3.133$, and
on a $6\times24^3$ lattice at $\beta=6.8$ corresponding
to $T/T_c=3.532$.  One further computation was made on an $8\times18^3$
lattice at $\beta=6.55$.  This corresponds to $T/T_c=2$ with lattice
spacing $a=1/8T$. The quark mass $m/T_c$ is kept independent of $a$
and $T$, and hence $ma$ decreases with increasing $T$ at fixed $N_t$ or 
with increasing $N_t$ at fixed $T$ according to eq.\ (\ref{mass}).

The lattice scale has been set using the plaquette measurements of
\cite{bielef} and the analysis performed in \cite{myold}. The conversion
of $\beta$ to $T$ using data for $N_t=4$ and 8 give the two curves in
Figure \ref{fg.temp}. The difference between the two curves is
within the statistical uncertainty of 5\% on the scale assignment.
There is the same degree of statistical uncertainty in the bare quark
masses and in all other scales. We shall not display this inherent scale
uncertainty in the measurements, but it should be kept in mind.

The simulations have been performed with a Cabbibo-Marinari
pseudo-heatbath technique with 3 $SU(2)$ subgroups updated on each
hit. An initial 1000 sweeps have been discarded for thermalisation.
On the $N_t\le6$ lattices we have used 80 configurations separated by
1000 sweeps for the measurements. On the $N_t=8$ lattice we have used
55 configurations separated by 500 sweeps.

\begin{figure}[hbt]\begin{center}
   \includegraphics{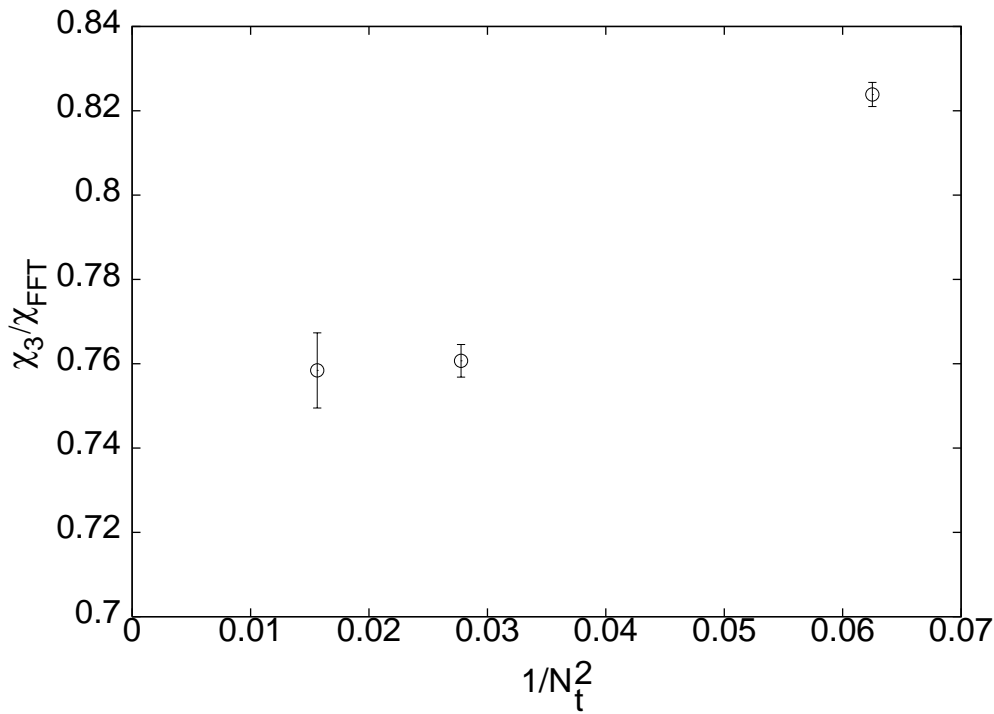}
   \end{center}
   \caption{$\chi_3/\chi^3_{FFT}$ at $T=2T_c$ shown as a function of
       $a^2\propto1/N_t^2$ for $m/T_c=0.03$. The ratio reaches its
       continuum limit when evaluated on lattices with $N_t=6$.}
\label{fg.extrap}\end{figure}

\begin{figure}[htb]\begin{center}
   \includegraphics{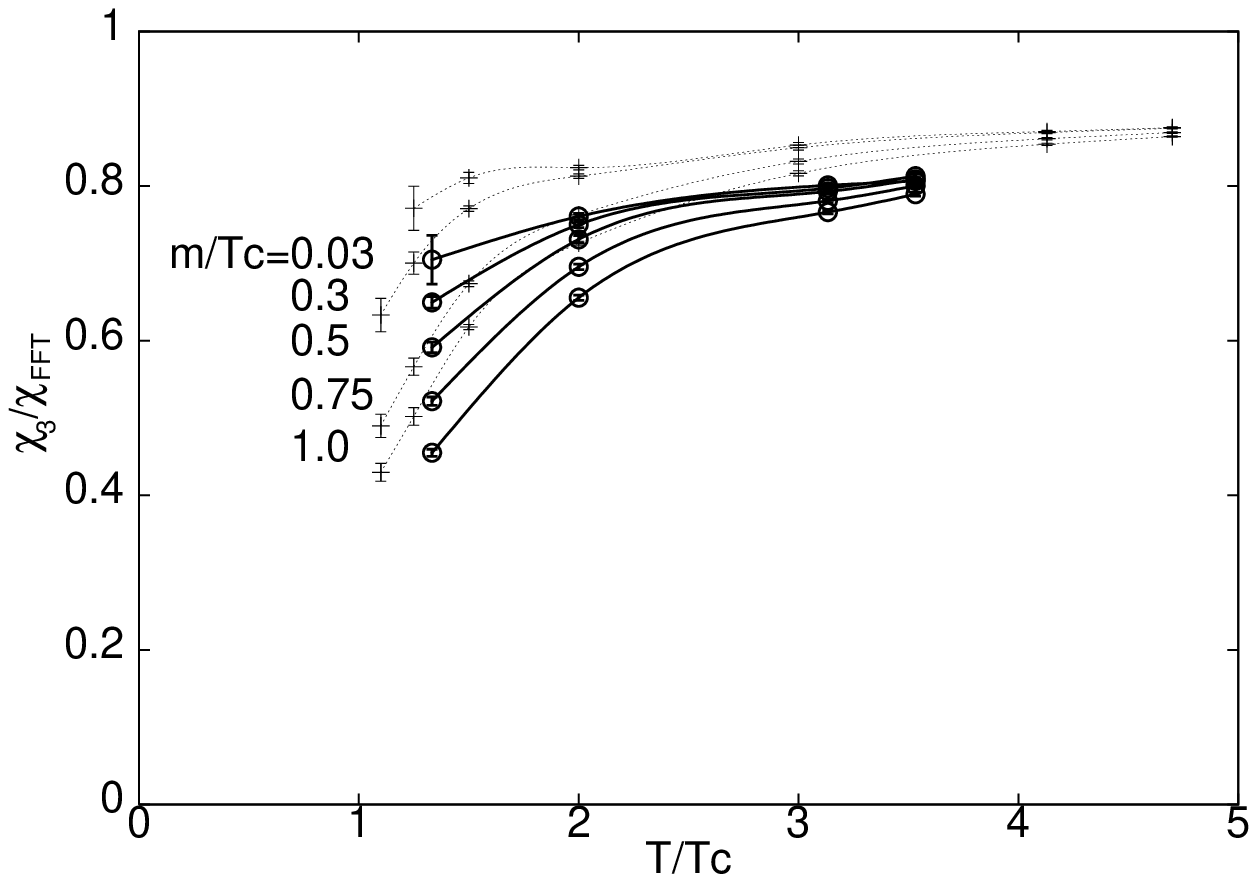}
   \end{center}
   \caption{The ratio $\chi_3/\chi^3_{FFT}$ in quenched QCD, shown as a
       function of $T$ for the values of $m/T_c$ indicated. Pluses denote
       data for lattice spacing $a=1/4T$ and circles for $a=1/6T$. The
       lines are cubic spline fits to the data.}
\label{fg.tdep}\end{figure}

We have previously checked that the spatial lattice size, $L=N_sa$,
is quite irrelevant to the values of $\chi$ measured as long as it is
sufficiently large compared to the inverse pion mass \cite{gavai,2fl}. In
fact, at all the couplings and quark masses we have used, $m_\pi
L\ge5$. The most important other constraint on $L$ comes from
the requirement that it should be large enough to prevent spatial
deconfinement \cite{saumen}.  This is ensured by taking $N_s/N_t\ge T/T_c$
in all our simulations.

The traces are evaluated by the usual stochastic technique,
\beq
   {\rm Tr}\,A = \frac1{2N}\sum_{i=i}^N R_i^\dag AR_i,
\label{stoch}\eeq
where $R_i$ are a set of $N$ uncorrelated vectors with components drawn
independently from a Gaussian ensemble with unit variance. Each vector
has three colour components at each site of the lattice. Since we use a
half lattice version of the Dirac operator for staggered Fermions, the
number of components of each vector $R_i$ is one and a half times the
number of lattice points.  $({\rm Tr}\,A)^2$ is obtained by dividing
the set of $R_i$ into disjoint blocks, constructing ${\rm Tr}\,A$
in each block, taking all possible distinct pairs of such estimates,
multiplying them and then averaging over the pairs.  For $N_t=4$ it is
possible to get accurate results with $N\approx10$, although we have
chosen to use $N=80$ for each gauge configuration \cite{2fl}.

It is easy to check that the variance of the estimator above is
proportional to ${\rm Tr}(A^2)$. Since the diagonal element of the
Fermion matrix is proportional to the quark mass $ma$, whereas the
off-diagonal elements are bounded by unity, with increasing $T$ or
decreasing $a$, ${\rm Tr}\,A$ decreases linearly, while its variance
remains constant. Hence, for sufficiently small $ma$ the number of
vectors has to increase quadratically with $ma$. We found that $N_v=100$
was sufficient for $N_t=8$ at $T=2T_c$, although at larger $N_t$ or $T$
significantly larger values of $N_v$ are needed. A further numerical
problem arises from the cancellation of the two matrix elements which
give $\chi_3$ (see eq.\ \ref{chi3}). While increasing $N_t$ and keeping
the physical size of the lattice fixed, each of these terms increases
quadratically but mutually cancel to give a number which decreases
quadratically with $N_t$. At fixed word length, this implies a reduction
in accuracy by a factor of $N_t^4$.

\begin{figure}[hbt]\begin{center}
   \includegraphics{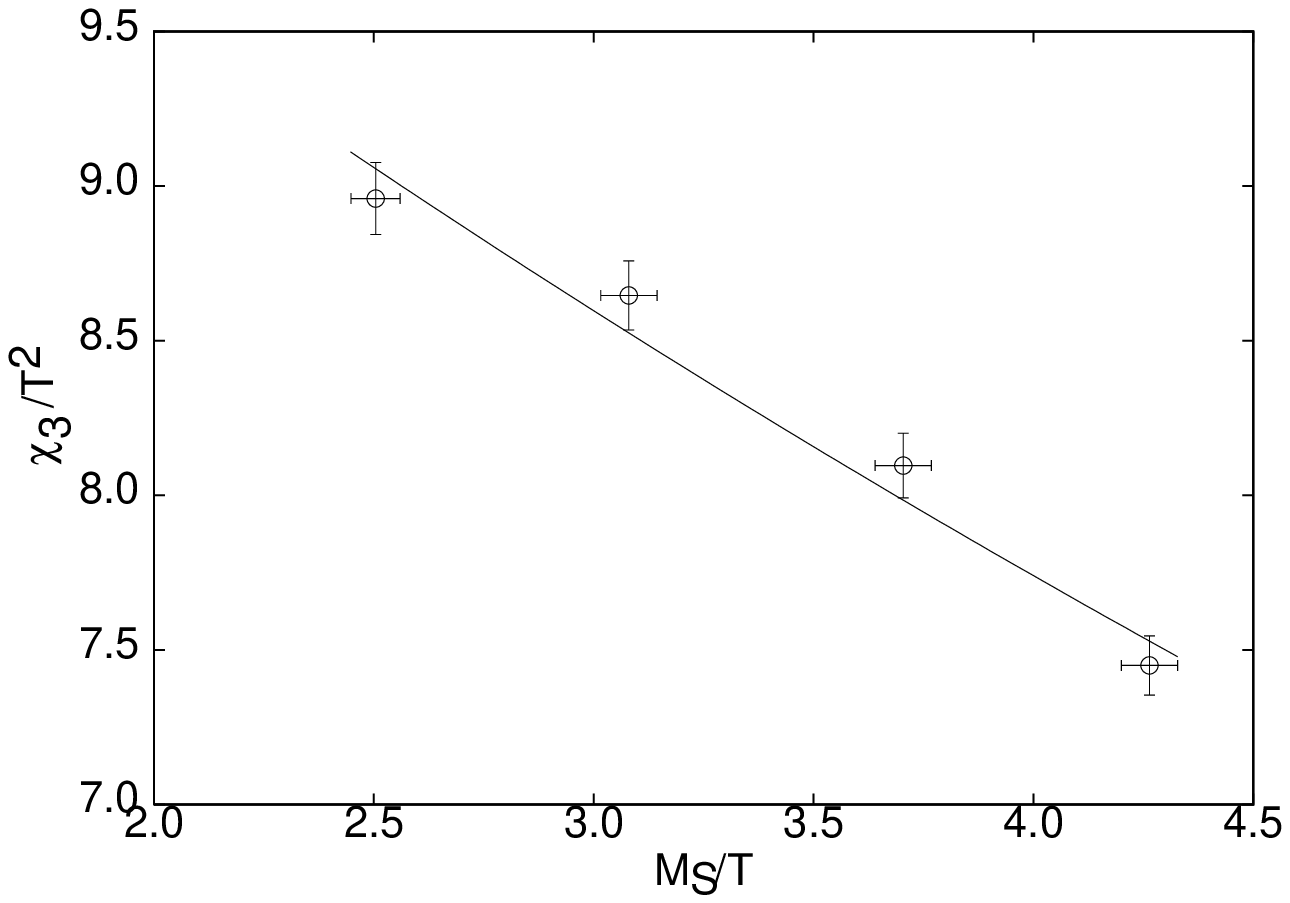}
   \end{center}
   \caption{Correlation between the screening mass in the scalar/pseudo-scalar
       sector, $M_S$, and the quark number susceptibility, $\chi_3$ at $2T_c$.
       The measurements have been made on a $8\times18^3$ lattice at a series
       of quark masses. The line is an exponential fit to the data.}
\label{fg.mpi}\end{figure}

\begin{table}\begin{center}\begin{tabular}{lrrrrr}
   \hline
   $T/T_c$ & $\qquad\chi_3/\chi^3_{FFT}$ & 
         $\qquad\qquad\chi_{ud}/T^2$ & $\qquad\chi_s/\chi^s_{FFT}$ &
         $\qquad\chi_0/\chi^0_{FFT}$ & $\qquad\chi_q/\chi^q_{FFT}$ \\
   \hline
   1.333 & 0.70  (3) & $(2\pm4)\times10^{-6}$ & 0.455 (5) &
           0.66  (2) & 0.68  (3) \\
   2.000 & 0.761 (3) & $(-1\pm2)\times10^{-7}$ & 0.656 (3) &
           0.740 (3) & 0.751 (3) \\
   3.133 & 0.801 (2) & $(-1\pm1)\times10^{-7}$ & 0.766 (2) &
           0.794 (2) & 0.798 (2) \\
   3.532 & 0.807 (2) & $(1\pm2)\times10^{-7}$ & 0.789 (2) &
           0.803 (2) & 0.805 (2) \\
   \hline
\end{tabular}\end{center}
\caption{Results for the continuum limit of quark number susceptibility
   in quenched QCD. We have taken $m_s/T_c=1$ as appropriate to full QCD.
   Our $N_t=4$ results indicate that at higher $T$ the ratios $\chi/\chi_{FFT}$
   remain flat.}
\label{tb.res}\end{table}

In Figure \ref{fg.extrap} we show how the continuum limit of $\chi_3$
is reached. There is an 8\% decrease in $\chi_3/\chi^3_{FFT}$ in going
from $N_t=4$ to $N_t=6$ lattices. However, within measurement errors,
the continuum limit is reached already at $N_t=6$. This implies a
cancellation of the $N_t$-dependence of $\chi_3$ and $\chi^3_{FFT}$. The
reason for the atypical behaviour for $N_t=4$ is not in the numerator,
which is non-perturbative, but in the denominator, which is the ideal gas.

For $N_t=4$, the spectrum of $p_0$ consists of $\{\pm\pi/4,
\pm3\pi/4\}$. As a result, $\sin^2p_0=1/2$ and the second term in
the sum on the right of eq.\ (\ref{fft}) vanishes identically. This
accident occurs only for $N_t=4$, and for all other values of $N_t$
this term contributes a non-vanishing value. It is easy to see that in
the limit $N_t\to\infty$ this term contributes to the leading part of
$\chi^3_{\scriptscriptstyle FFT}$.  This is the reason the values for
$\chi_3/\chi^3_{\scriptscriptstyle FFT}$ for $N_t=4$ lie well away from
the continuum limit.

Our complete results for the dependence of $\chi_3/\chi^3_{FFT}$
on $T/T_c$ at several different quark masses for $N_t=4$ and 6 are
shown in Figure \ref{fg.tdep}. In view of the data shown in Figure
\ref{fg.extrap}, our results for $N_t=6$ are the continuum limit for the
quenched theory: these are summarised in Table \ref{tb.res}.  Note that
even at temperature as high as $3.5T_c$ the result is far below the ideal
gas limit. Closer to $T_c$ there is an even more dramatic fall in the
value of $\chi_3/\chi^3_{FFT}$.  We have also measured the off-diagonal
susceptibility $\chi_{ud}$. As already seen for $N_t=4$ this is consistent
with zero. In Table \ref{tb.res} we show that the continuum limit of
$\chi_{ud}$ also vanishes within errors of $10^{-7}T^2$.  It is therefore
much smaller than the 3-loop perturbative result given in \cite{toni}.

Note that the continuum limit of our measurements lie 20--30\% below
the ideal gas result. They also differ significantly from HTL as well
as the skeleton graph resummed results of \cite{toni}. At $T=3T_c$,
HTL predicts $\chi_3/\chi^3_{FFT}=0.90$--$0.94$ on varying the
scale of $\alpha_s$ between $\pi T$ and $4\pi T$, and the resummed
computation gives $\chi_3/\chi^3_{FFT}=0.95$--$0.97$. Our measurement
shows $\chi_3/\chi^3_{FFT}=0.80$ at $3T_c$ (see Table \ref{tb.res}).
A second computation of the HTL result is available for $N_f=2$
\cite{mustafa}. Since this agrees with the HTL result of \cite{toni}
for $N_f=2$, we believe the HTL result for $\chi_3/\chi^3_{FFT}$ is
under control. There is thus a genuine discrepancy between these lattice
results and existing perturbative computations.

The quark number susceptibilities are closely related to the screening
correlator of a one-link separated quark bilinear operator which
corresponds to a $\rho$ meson at zero temperature. The $T>0$ transfer
matrix of the problem mixes this with a quark bilinear that corresponds
to the $\pi$ \cite{old}. We have earlier shown evidence for $N_t=4$
that $\chi_3$ is closely related to the only known non-perturbative
quantity among the screening correlators, that is the screening mass $M_S$
coming from the two degenerate correlators that descend from the $\pi$
and $\sigma$ operators of the zero temperature theory \cite{2fl}. In
view of the strongly non-perturbative character of $\chi_3$, as revealed
by the comparison with HTL and resummation explained above, examining
this correlation in the continuum becomes more significant. In Figure
\ref{fg.mpi} we show data obtained on $8\times18^3$ lattices which
indicate that this correlation survives into the continuum.

\section{Strangeness}

\begin{figure}[hbt]\begin{center}
   \includegraphics{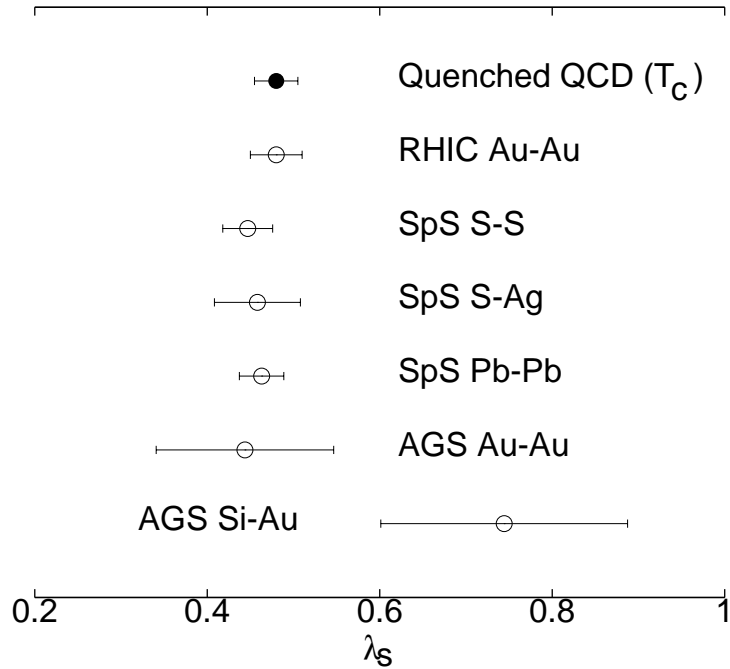}
   \end{center}
   \caption{The Wroblewski parameter, $\lambda_s$, obtained from various
       experimental measurements, compared with the value obtained from
       the lattice measurements reported here.}
\label{fg.wrob}\end{figure}

Measurements of hadron yields in relativistic heavy-ion collisions
have been analysed extensively \cite{heidel,strange} and the observed
enhancement of strangeness has been claimed to be evidence for quark-gluon
plasma formation. The chemical composition is observed at the hadron
freezeout temperature, which has been found to be very close to the
value of $T_c$ in full QCD. It has been argued that this chemical
composition cannot arise due to hadronic rescattering \cite{heinz}.
It would be interesting if it could be directly determined whether or not
the composition is characteristic of the plasma for $T\ge T_c$. We argue
below that this question can be answered by a lattice QCD measurement
such as we have presented above.

The quantity which seems to be directly connected to lattice measurements
is the Wroblewski parameter \cite{wrob}, $\lambda_s$, which measures
the ratio of newly created primary strange to light quarks---
\beq
   \lambda_s = \frac{2\langle s\bar s\rangle}{\langle u\bar u+d\bar d\rangle}.
\label{wrob}\eeq
A fluctuation-dissipation theorem relates the rate of production of
quark pairs by the plasma in equilibrium to the imaginary part of
the full complex quark number susceptibilities \cite{fw}. If the inverse of
the characteristic time scales of the QCD plasma are not close to the
typical energy scales for the production of strange and lighter quarks,
then the ratio of their production rates is just the ratio of the static
susceptibilities that we have measured.  Then, if the observed chemical
composition is created in equilibrium, we should have
\beq
   \lambda_s = \frac{2\chi_s}{\chi_u+\chi_d} \approx \frac{\chi_s}{\chi_u},
\label{relwrob}\eeq
where the last equality holds in the limit of equal $u$ and $d$ quark
masses. These susceptibilities have to be evaluated on the lattice at the
temperature and chemical potential, $\mu_0$, relevant to the collision.
However, it has been shown recently \cite{cleymans} that $\lambda_s$ is
insensitive to $\mu_0$ for SpS and RHIC energies. Furthermore a lattice
measurement shows that the screening mass, $M_S$, does not change rapidly
with increasing chemical potential \cite{miyamura}. Then, in view of
the correlation between $M_S$ and $\chi$ shown in Figure \ref{fg.mpi},
we do not expect rapid changes with chemical potential in the ratio
above. Both these arguments show that as a first approximation one can
take the ratio in eq.\ (\ref{relwrob}) at zero chemical potential.

From our measurements of $\chi$ reported here, we can form the lattice
``prediction'' of $\lambda_s$ created in the plasma at $T_c$. This
has been done by taking our measurements at $T/T_c=4/3$ and 2, and
extrapolating them linearly to $T_c$. For $\chi_u$ we use data taken
for $m/T_c=0.03$, which is light enough for this purpose. Since scaled
quantities feel the smallest correction for unquenching \cite{2fl},
we take $\chi_s$ to correspond to $m/T_c=1$, which is appropriate to
full QCD.  In Figure \ref{fg.wrob} the resultant prediction is compared
with the data collected in \cite{strange,cleymans}.  It can be seen that
there is a fair agreement with the data.  However, this is subject to
several assumptions---
\begin{enumerate}
\item The foremost assumption is that the characteristic time scales of the
   plasma are not close to the inverse energy scale of the production
   processes. It is not possible to test this assumption in an Euclidean
   computation. However, it has been suggested that these characteristic
   time scales could be observed in real or virtual photon production as
   spikes in the spectrum \cite{spikes}. If these are seen and they lie in
   the energy range for the production, the assumption would be falsified.
\item Another important assumption is that chemical equilibration takes place
   in the plasma. If chemical equilibrium is not achieved but the light
   degrees of freedom achieve energy equipartition, then the effective
   temperature could be higher.
\item We have assumed slow variation of the ratio $\chi_s/\chi_u$ with
   chemical potential. While this seems to be a reasonable estimate in view
   of the results of \cite{cleymans,miyamura} and our own observations about
   the relation between the lowest screening mass and the susceptibilities,
   one should keep in mind possible changes in the ratio due to violation of
   this assumption.
\item Our continuum results are obtained in quenched QCD, whereas the data
   corresponds to full QCD. As mentioned already, we have taken $m/T_c=1$,
   as appropriate to full QCD, in order to correct for this. However, an
   additional 5--10\% shift in our results could arise when unquenching
   \cite{2fl}. This is not shown in Figure \ref{fg.wrob}.
\item Our measurements are made away from $T_c$ and extrapolated down to
   this temperature. In principle this can be corrected by a computation
   made directly at $T_c$. However, due to the enormous cost of extracting
   physical quantities at $T_c$, we postpone this to the time when we
   attempt a full QCD simulation.
\end{enumerate}

\noindent
It might seem surprising that $\lambda_s$ corresponds to the chemical
composition at $T_c$. However, a moment's thought shows that this is
entirely in accord with the assumption of chemical equilibrium. Even
if the system is thermalized at a much higher temperature, it remains
in equilibrium until it reaches $T_c$. Departure from equilibrium as
the system cools through the transition is unlikely, and if it occurs,
should have a visible signature such as the formation of a disoriented
chiral condensate \cite{rajag}. In the absence of such phenomena, if
$\lambda_s$ does not correspond to the chemistry at $T_c$, then the
second assumption above would be demonstrated to be false. Also, with
increasing ion energy the system moves closer to zero baryon number, as
seen at RHIC. This makes our third assumption better at the RHIC and LHC.
The fourth assumption can be removed by a computation with full QCD. Such
a computation is planned, where also the final assumption can be removed,
and will be reported elsewhere.

\section{Summary}

In summary, we have presented new and precise results on quark number
susceptibilities over a wide range of temperatures and quark masses
in the high temperature phase of QCD. The main aim of this study was
to obtain the continuum extrapolation of these susceptibilities. As
shown in Figure \ref{fg.extrap}, while there is a significant change
in going from $N_t=4$ to $N_t=6$ lattices, there is no statistically
significant change in the susceptibilities beyond this. The results
obtained for $N_t=6$ lattices can then be taken as the continuum limit.
This limit is shown in Figure \ref{fg.tdep} and Table \ref{tb.res}.

There is a strong discrepancy between the continuum extrapolated lattice
results and HTL computations for these quantities--- varying between
about 12\% at the highest $T$ to even larger at smaller $T$. The
off-diagonal susceptibility, $\chi_{ud}$, is zero, in contrast to
the HTL results. There is a somewhat larger discrepancy between the
continuum extrapolated lattice results and the skeleton graph resummed
results. The conjecture that there is a strong non-perturbative component
to the quark number susceptibilities is supported by an observed strong
correlation between the smallest quark bilinear screening mass, $M_S$,
and the susceptibility $\chi_3$ (Figure \ref{fg.mpi}).

It is interesting to note that the continuum extrapolated results for
the strange and light quark susceptibilities can be used to give a
surprisingly good description of the chemical composition of hadrons at
freezeout in SpS and RHIC experiments (Figure \ref{fg.wrob}). This has to
be treated as a preliminary estimate due to the many caveats which we have
listed, and some of which we plan to check in future lattice measurements.


\begin{thebibliography}{99}
\bibitem{gott}
  S.\ Gottlieb {\sl et al.\/}, {\sl Phys.\ Rev.\ Lett.\/} 59 (1987) 2247.
\bibitem{gavai}
  R.\ V.\ Gavai {\sl et al.\/}, {\sl Phys.\ Rev.\/} D 40 (1989) 2743;\\
  C.\ Bernard {\sl et al.\/}, {\sl Phys.\ Rev.\/} D 54 (1996) 4585;\\
  S.\ Gottlieb {\sl et al.\/}, {\sl Phys.\ Rev.\/}, D 55 (1997) 6852.
\bibitem{qnch}
  R.\ V.\ Gavai and S.\ Gupta, {\sl Phys.\ Rev.\/} D 64 (2001) 074506.
\bibitem{2fl}
  R.\ V.\ Gavai, S.\ Gupta and P.\ Majumdar, hep-lat/0110032,
    {\sl Phys.\ Rev.\/} D, in press.
\bibitem{koch}
   M.\ Asakawa, U.\ Heinz and B.\ M\"uller, {\sl Phys.\ Rev.\ Lett.\/} 85 (2000) 2072;\\
   S.\ Jeon and V.\ Koch, {\sl ibid.\/} 85 (2000) 2076;\\
   D.\ Bower and S.\ Gavin, {\sl Phys.\ Rev.\/} C 64 (2001) 051902;\\
   S.\ Jeon, V.\ Koch, K.\ Redlich and X.\ N.\ Wang, nucl-th/0105035.
\bibitem{toni}
   J.-P.\ Blaizot, E.\ Iancu and A.\ Rebhan, hep-ph/0110369.
\bibitem{saumen}
  S.\ Datta and S.\ Gupta, {\sl Phys.\ Lett.\/}, B 471 (2000) 382.
\bibitem{bielef}
  G.\ Boyd {\sl et al.\/}, {\sl Nucl.\ Phys.\/}, B 469 (1996) 419.
\bibitem{myold}
  S.\ Gupta, {\sl Phys.\ Rev.\/}, D 64 (2001) 034507.
\bibitem{mustafa}
  P.\ Chakraborty, M.\ G.\ Mustafa and M.\ H.\ Thoma, hep-ph/0111022,
    to appear in {\sl Euro.\ Phys.\ J.\/} C.
\bibitem{old}
  S.\ Gupta, {\sl Phys.\ Rev.\/}, D 60 (1999) 094505.
\bibitem{heidel}
  P.\ Braun-Munzinger {\sl et al.\/}, {\sl Phys. Lett.\/}, B 465 (1999) 43;\\
  J.\ Letessier and J.\ Rafelski, {\sl Nucl.\ Phys.\/}, A 661 (1999) 97c.
\bibitem{strange}
  F.\ Becattini {\sl et al.\/}, {\sl Phys.\ Rev.\/} C 64 (2001) 024901.
\bibitem{heinz}
  U.\ Heinz {\sl J.\ Phys.\/} G 25 (1999) 263;\\
  R.\ Stock, {\sl Phys.\ Lett.\/} B 456 (1999) 277.
\bibitem{wrob}
  A.\ Wroblewski, {\sl Acta Phys.\ Pol.\/}, B 16 (1985) 379.
\bibitem{fw}
  P.\ C.\ Martin, in {\sl Many Body Physics\/}, Proceedings of the 1967 Les
    Houches School, Eds.\ C.\ DeWitt and R.\ Balian, Gordon and Breach,
    New York, 1968;\\
  W.\ Marshall and S.\ W.\ Lovesey, {\sl Theory of Thermal Neutron
    Scattering\/}, Oxford University Press, London, 1971;\\
  A.\ L.\ Fetter and J.\ D.\ Walecka, {\sl Quantum Theory of Many-particle
     Systems\/}, McGraw-Hill, New York, 1971.
\bibitem{cleymans}
  J.\ Cleymans, e-print hep-ph/0201142.
\bibitem{miyamura}
  S.\ Choe {\sl et al.\/}, e-print hep-lat/0110223.
\bibitem{spikes}
  H.\ A.\ Weldon, {\sl Nucl.\ Phys.\/} A 525 (1991) 405c;\\
  S.\ Gupta, {\sl Nucl.\ Phys.\/} A 566 (1994) 69c.
\bibitem{rajag}
  K.\  Rajagopal and F.\ Wilczek, {\sl Nucl.\ Phys.\/}, B 404 (1993) 577.
\end{thebibliography}
\end{document}